\documentstyle[12pt]{article}
\def\journal#1, #2, #3, #4 { {\sl #1~}{\bf #2~}(#3) #4 }

\def\cmp{\journal Comm. Math. Phys., }

\def\np{\journal Nucl. Phys., }

\def\pl{\journal Phys. Lett., }

\catcode`\@=11
\def\marginnote#1{}
\newcount\hour
\newcount\minute
\newtoks\amorpm
\hour=\time\divide\hour by60
\minute=\time{\multiply\hour by60 \global\advance\minute
by-\hour}\edef\standardtime{{\ifnum\hour<12
\global\amorpm={am}%
        \else\global\amorpm={pm}\advance\hour by-12 \fi
        \ifnum\hour=0 \hour=12 \fi
        \number\hour:\ifnum\minute<10
0\fi\number\minute\the\amorpm}}
\edef\militarytime{\number\hour:\ifnum\minute<10
0\fi\number\minute}

\def\draftlabel#1{{\@bsphack\if@filesw {\let\thepage\relax
   \xdef\@gtempa{\write\@auxout{\string
      \newlabel{#1}{{\@currentlabel}{\thepage}}}}}\@gtempa
   \if@nobreak \ifvmode\nobreak\fi\fi\fi\@esphack}
        \gdef\@eqnlabel{#1}}
\def\@eqnlabel{}
\def\@vacuum{}
\def\draftmarginnote#1{\marginpar{\raggedright\scriptsize\tt#1}}
\def\draft{\oddsidemargin -.5truein
        \def\@oddfoot{\sl preliminary draft \hfil
        \rm\thepage\hfil\sl\today\quad\militarytime}
        \let\@evenfoot\@oddfoot \overfullrule 3pt
        \let\label=\draftlabel
        \let\marginnote=\draftmarginnote

\def\@eqnnum{(\theequation)\rlap{\kern\marginparsep\tt\@eqnlabel}%
\global\let\@eqnlabel\@vacuum}  }

\def\numberbysection{\@addtoreset{equation}{section}
        \def\theequation{\thesection.\arabic{equation}}}

\def\underline#1{\relax\ifmmode\@@underline#1\else
 $\@@underline{\hbox{#1}}$\relax\fi}

\catcode`@=12
\relax

\def\fin{\end{document}}
\def\beq{\begin{equation}}
\def\eeq{\end{equation}}
\def\beqa{\begin{eqnarray}}
\def\eeqa{\end{eqnarray}}
 \def\nnn{\nonumber \\}
\def\sqr#1#2{{\vcenter{\vbox{\hrule height.#2pt
\hbox{\vrule width.#2pt height#1pt \kern#1pt
\vrule width.#2pt}
\hrule height.#2pt}}}}

\def\Jne#1 {J_{#1}^e\, \!}
\def\Jneb#1 {{\overline J}_{#1}^e\, \!}
\def\Jnep#1 {J_{#1}'\, \!^e\, \!}
\def\Jnebp#1 {{\overline J}_{#1}'\, \!  \!^e\, \!}

\def\hhat{{\widehat h}}

\def\Jhat{{\widehat J}}

\def\mhat{{\widehat m}}

\def\chib{{\overline \chi}}

\def\qhat{{\widehat q}}

\def\mhat{{\widehat m}}

\def\Shat{{\widehat S}}

\def\Jb{{\overline J}}

\def\Sb{\overline S}
\def\mb{{\overline  m}}
\def\mub{{\bar \mu}}
\def\zb{{\bar z}}

\def\Vb{{\overline V}}

\def\varthetab{\overline \vartheta}
\def\mhatb{\widehat {\overline S}}

\def\Vt{{\widetilde V}}
\def\Vtb{{\widetilde {\overline V}}}
\def\pb{\overline p}

\def\Jhatb{\widehat {\overline J}}
\def\mhatb{\widehat {\overline m}}
\def\Shatb{\widehat {\overline S}}

\def\Jgen#1 {  {\underline J_{#1}} }
\def\Jgenp#1 #2 {(J_{#1}+{#2},\Jhat_{#1})}
\def\Jgenm#1 #2 {(J_{#1}-{#2},\Jhat_{#1})}
\def\Jg#1 {J_{#1},\Jhat_{#1}}
\def\Jgp#1 #2 {J_{#1}+{#2},\Jhat_{#1}}
\def\Mgen#1 {{\underline M_{#1}}}

\def\produit#1,#2,#3,#4 {P\Bigl (
[{#1},{#2}]\otimes\{{#3}\},{#4}\Bigr )}
\def\produitscript#1,#2,#3,#4 {P\Bigl (
[{\scriptstyle{#1},{#2}}]\otimes\{{\scriptstyle{#3}}\},{#4}\Bigr
)}
\def\pprod#1,#2,#3,#4,#5 {P\Bigl (
[{#1},{#2}]\otimes[{#3},{#4}],{#5}\Bigr )}
\def\pprodscript#1,#2,#3,#4,#5 {P\Bigl (
[{\scriptstyle{#1},{#2}}]\otimes[{\scriptstyle{#3},{#4}}],{#5}\Bigr
)}

\def\fusV#1,#2,#3,#4,#5,#6 {f_V(
\Jgen{#1} ,
\Jgen{#2} ,
\Jgen{#3} ,
\Jgen{#4} ,
\Jgen{#5} ,
\Jgen{#6} )}

\def\brdV#1,#2,#3,#4,#5,#6 {b_V(
\Jgen{#1} ,
\Jgen{#2} ,
\Jgen{#3} ,
\Jgen{#4} ,
\Jgen{#5} ,
\Jgen{#6} )}

\def\fusxi#1,#2,#3 {f_\xi (\Jgen{#1} ,
\Mgen{#1} ,
\Jgen{#2} ,
\Mgen{#2} ,
\Jgen{#3} )}

\def\gaghat{{\hat {\bigl \{}}}
\def\gadhat{{\hat {\bigr \}}}}

\def\sixjxi#1,#2,#3,#4,#5,#6 {{\left\{\left . \!\! \,^{#1}_{#2}
\,^{#3}_{#4} \right | \!\, ^{#5}_{#6}\right\}}}
\def\sixje#1,#2,#3,#4,#5,#6 {{\left\{\left\{\left . \!\! \,
^{#1}_{#2}
\, ^{#3}_{#4} \right | \!\, ^{#5}_{#6}\right\}\right\}}}
\def\sixjxihat#1,#2,#3,#4,#5,#6 {{{\gaghat\left . \!\! \,
^{#1}_{#2}
\, ^{#3}_{#4} \right | \!\, ^{#5}_{#6}\gadhat}}}
\def\sixjehat#1,#2,#3,#4,#5,#6 {{\gaghat\gaghat\left . \!\! \,
^{#1}_{#2}
\, ^{#3}_{#4} \right | \!\, ^{#5}_{#6}\gadhat\gadhat}}

\def\pb{\overline p}

\def\Jhatb{\widehat {\overline J}}
\def\mhatb{\widehat {\overline m}}
\def\vertex#1,#2,{{{\cal V}^{{#1,#2}}}}
\def\vertexc#1,#2,{{{\cal V}^{{#1,#2}}_{conj}}}
\def\lf#1,#2,{{L_{#1,#2}}}
\def\lfc#1,#2,{{{\bar L}_{#1,#2}}}

\def\abstracts#1{{

\centering{\begin{minipage}{12.2truecm}\footnotesize\baselineskip=12pt\noindent
        \centerline{\footnotesize ABSTRACT}\vspace*{0.3cm}
        \parindent=0pt #1
        \end{minipage}}\par}}
\catcode`\@=11
\long\def\@makefntext#1{
\protect\noindent \hbox to 3.2pt {\hskip-.9pt
$^{{\ninerm\@thefnmark}}$\hfil}#1\hfill}                %CAN BE USED

\def\@makefnmark{\hbox to 0pt{$^{\@thefnmark}$\hss}}  %ORIGINAL

\def\ps@myheadings{\let\@mkboth\@gobbletwo
\def\@oddhead{\hbox{}
\rightmark\hfil\ninerm\thepage}
\def\@oddfoot{}\def\@evenhead{\ninerm\thepage\hfil
\leftmark\hbox{}}\def\@evenfoot{}
\def\sectionmark##1{}\def\subsectionmark##1{}}

\renewcommand{\thefootnote}{\fnsymbol{footnote}}

\newcounter{sectionc}\newcounter{subsectionc}\newcounter{subsubsectionc}
\renewcommand{\section}[1] {\vspace*{0.6cm}\addtocounter{sectionc}{1}
\setcounter{subsectionc}{0}\setcounter{subsubsectionc}{0}\noindent
        {\normalsize\bf\thesectionc. #1}\par\vspace*{0.4cm}}
\renewcommand{\subsection}[1] {\vspace*{0.6cm}\addtocounter{subsectionc}{1}
        \setcounter{subsubsectionc}{0}\noindent
        {\normalsize\it\thesectionc.\thesubsectionc. #1}\par\vspace*{0.4cm}}
\renewcommand{\subsubsection}[1]
{\vspace*{0.6cm}\addtocounter{subsubsectionc}{1}
        \noindent
{\normalsize\rm\thesectionc.\thesubsectionc.\thesubsubsectionc.
        #1}\par\vspace*{0.4cm}}

\newcounter{appendixc}
\newcounter{subappendixc}[appendixc]
\newcounter{subsubappendixc}[subappendixc]

\renewcommand{\appendix}[1] {\vspace*{0.6cm}
        \refstepcounter{appendixc}
        \setcounter{figure}{0}
        \setcounter{table}{0}
        \setcounter{equation}{0}
        \renewcommand{\thefigure}{\Alph{appendixc}.\arabic{figure}}
        \renewcommand{\thetable}{\Alph{appendixc}.\arabic{table}}
        \renewcommand{\theappendixc}{\Alph{appendixc}}
        \renewcommand{\theequation}{\Alph{appendixc}.\arabic{equation}}
        \noindent{\bf Appendix \theappendixc #1}\par\vspace*{0.4cm}}

\def\abstracts#1{{

\centering{\begin{minipage}{12.2truecm}\footnotesize\baselineskip=12pt\noindent
        \centerline{\footnotesize ABSTRACT}\vspace*{0.3cm}
        \parindent=0pt #1
        \end{minipage}}\par}}

\renewenvironment{thebibliography}[1]
        {\begin{list}{\arabic{enumi}.}
        {\usecounter{enumi}\setlength{\parsep}{0pt}
\setlength{\leftmargin 1.25cm}{\rightmargin 0pt}
         \setlength{\itemsep}{0pt} \settowidth
        {\labelwidth}{#1.}\sloppy}}{\end{list}}

\newcounter{itemlistc}
\newcounter{romanlistc}
\newcounter{alphlistc}
\newcounter{arabiclistc}

\newcommand{\fcaption}[1]{
        \refstepcounter{figure}
        \setbox\@tempboxa = \hbox{\footnotesize Fig.~\thefigure. #1}
        \ifdim \wd\@tempboxa > 6in
           {\begin{center}
        \parbox{6in}{\footnotesize\baselineskip=12pt Fig.~\thefigure. #1}
            \end{center}}
        \else
             {\begin{center}
             {\footnotesize Fig.~\thefigure. #1}
              \end{center}}
        \fi}

\newcommand{\tcaption}[1]{
        \refstepcounter{table}
        \setbox\@tempboxa = \hbox{\footnotesize Table~\thetable. #1}
        \ifdim \wd\@tempboxa > 6in
           {\begin{center}
        \parbox{6in}{\footnotesize\baselineskip=12pt Table~\thetable. #1}
            \end{center}}
        \else
             {\begin{center}
             {\footnotesize Table~\thetable. #1}
              \end{center}}
        \fi}

\def\@citex[#1]#2{\if@filesw\immediate\write\@auxout
        {\string\citation{#2}}\fi
\def\@citea{}\@cite{\@for\@citeb:=#2\do
        {\@citea\def\@citea{,}\@ifundefined
        {b@\@citeb}{{\bf ?}\@warning
        {Citation `\@citeb' on page \thepage \space undefined}}
        {\csname b@\@citeb\endcsname}}}{#1}}

\newif\if@cghi
\def\cite{\@cghitrue\@ifnextchar [{\@tempswatrue
        \@citex}{\@tempswafalse\@citex[]}}
\def\citelow{\@cghifalse\@ifnextchar [{\@tempswatrue
        \@citex}{\@tempswafalse\@citex[]}}
\def\@cite#1#2{{$\null^{#1}$\if@tempswa\typeout
        {IJCGA warning: optional citation argument
        ignored: `#2'} \fi}}

 1
 1
 1

\font\ninerm=cmr9

\begin{document}
\begin{flushright}
LPTENS 95/25\\
hep-th/9506040
\end{flushright}

\centerline{\normalsize\bf THE NEW PHYSICS OF}
\baselineskip=22pt
\centerline{\normalsize\bf STRONGLY COUPLED 2D GRAVITY}

\centerline{\sl (Strings95, USC,
Los Angeles March 13-18).}
\baselineskip=16pt

%\vfill
%\vspace*{0.6cm}
\centerline{\footnotesize Jean-Loup GERVAIS}
\baselineskip=13pt
\centerline{\footnotesize\it  Laboratoire de Physique Th\'eorique
}
\baselineskip=12pt
\centerline{\footnotesize\it  de l'\'Ecole Normale Sup\'erieure,}
\centerline{\footnotesize\it 24 rue Lhomond,
75231 Paris CEDEX 05, ~France}
\centerline{\footnotesize E-mail:gervais@physique.ens.fr}

%\vfill
\vspace*{0.9cm}
\abstracts{
The strong coupling physics of two dimensional gravity
at $C=7$, $13$, $19$ is
summarized. It is based on a new set of local fields which
do not preserve chirality. Thus this quantum number becomes
``deconfined'' in the strongly coupled regime. This new set
leads to  a novel
definition of the area elements, and hence
to a modified expression for the  string suceptibility, which is
the real part of  the KPZ formula. It allows to define
topological (strongly coupled )
Liouville string theories (without transverse degree
of freedom)  which are completely solvable, and are
natural extensions of
 the standard matrix  models.
 }

%\vspace*{0.6cm}
\normalsize\baselineskip=15pt
\setcounter{footnote}{0}
\renewcommand{\thefootnote}{\alph{footnote}}

\section{Introduction}
The point of this seminar is to show, following refs.\cite{G3,GR1,GR3,GR4} how
the operator approach to
Liouville theory remains
applicable to the strong coupling regime.
Although the Liouville exponentials
loose meaning in the strongly coupled regime (since their
operator product algebra involves operators and/or
highest weight
states with
complex  Virasoro weights),  the general
operator-family of their chiral components may still be
used, when
truncation theorems\cite{GN5,G3,GR1} apply, that is with central charges $C=7$,
$13$, $19$.
Indeed for these values
 there exist
subfamilies of the  above chiral operators
which form  closed operator algebras, and
are compatible with the reality condition of  Virasoro weights.
In the present operator approach they   are  used to construct
a new set of local fields which replace the Liouville
exponentials. Since  both sets are constructed out of the same
free B\"acklund fields, they may be considered as  related by
a new type of quantum B\"acklund transformation, that connect the
weak and strong coupling regimes of two-dimensional gravity.
 In the present lecture notes,
I summarise   the basic  features (deconfinement of chirality,
new expression for the string susceptibility)  of these
new set of local fields that replace the Liouville exponentials,
in the strong coupling regime. Moreover, the basic properties of the
new topological models will be recalled, where the gravity part is in the
strong coupling regime. The message will be that
the derivation of the new features is very close to
the previous weak-coupling  one, once the new set of local
physical fields is established.

\section{The weak coupling regime revisited}
Remarkably, the present operator method treats the weak and
strong coupling regimes  much on the same footing.
As a preparation, let us  recall some basic points  of
the weak-coupling discussion\cite{G5,GS1,CGR}.
Our conventions are to let
\beq
 Q_L=\sqrt{(C-1)/3}, \quad Q_M=\sqrt{(25-C)/3}
\label{Qdef}
\eeq
where $C$ is the Liouville central charge. We call $\vartheta_L$,
and $\varthetab_L$ the two chiral components of the
B\"acklund free field associated with the
Liouville field $\Phi$. The building blocks  of the quantum group
approach to Liouville theory are chiral fields of the form
\beq
\Vt^{(J\, \Jhat)}_{m \mhat}\propto V^{(J\, \Jhat)}_{-J \, -\Jhat}\>
S^{J+m}\,
\Shat^{\Jhat+\mhat}, \quad
\Vtb^{(\Jb\, \Jhatb)}_{\mb \, \mhatb}\propto
\Vb^{(\Jb\, \Jhatb)}_{-\Jb\,
-\Jhatb}\>
\Sb^{\Jb+\mb}\,
\Shatb^{\Jhatb+\mhatb},
\label{Vgendef}
\eeq
The $V$ fields are functions of $z$, and the $\Vb$ fields
functions of $\zb$. The fields $V_{-J -\Jhat}^{(J\Jhat)}$ are
simple exponentials
which may be directly re-expressed in terms of the  B\"acklund
free
fields.
\beq
V_{-J -\Jhat}^{(J\Jhat)}=\ :\exp\left [(J\alpha_-+\Jhat
\alpha_+)\vartheta_L\right ]:\ ,\quad
\Vb^{(\Jb\, \Jhatb)}_{-\Jb\,
-\Jhatb}=\ :\exp\left [(\Jb \alpha_-+\Jhatb
\alpha_+)\varthetab_L\right ]:\ .
\label{VJJhatdef}
\eeq
$S$, $\Sb$,  and $\Shat$, $\Shatb$
 are the screening operators\cite{GS1}
associated with the two screening
charges
\beq
\alpha_\pm={Q_L/ 2}\pm
 i {Q_M/ 2}.
\label{alphadef}
\eeq
The $J$'s are quantum group spins, but
we will not dwell upon this aspect. In practice they determine
the weight of the $V$ operators which are of the
type $(2\Jhat+1,2J+1)$ in the BPZ classification.
We deal with irrational
theories, since this will be the case in the strong coupling
regime. Thus the range of $J$ and $\Jhat$ is unbounded.
Concerning the Hilbert space,
the Verma modules is of course  charaterized by the same
four quantum group
spins. The highest weight states will be noted
$|J,\, \Jhat> |\Jb,\, \Jhatb>$.
 Following  the earlier studies\cite{CGR},   the
chiral operators just written  are noted with a tilde to
emphasize their special normalization which is such that their
fusing and braiding matrices are exactly equal to q-6j  symbols.
The corresponding highest-weight  matrix elements
of the chiral fields
define the coupling constants\cite{CGR,GR1}.

The local
Liouville exponentials are given by\cite{G3,GS1}
\beq
e^{\textstyle -(J\alpha_-+\Jhat\alpha_+)\Phi(z, \zb )}=
\sum _{m,\, \mhat}
\Vt_{m\,\mhat}^{(J,\,\Jhat)}(z)\,
\Vtb_{m\,\mhat}^{(J,\,\Jhat)}(\zb) \>
\label{Lexpdef}
\eeq
This form is dictated by locality and closure  under  fusion. For the
following, it is important to stress that this expression only
involves $V$ and $\Vb$ fields  with equal quantum numbers
($J=\Jb$, $m=\mb$, and so on), while $m$, amd $\mhat$ are summed
over independently. Thus the Liouville exponentials have zero
conformal spins. On the other hand, it follows from the formulae
just summarized that
\beq
<J_2,\, \Jhat_2 |  \Vt^{(J\Jhat)}_{m \mhat} | J_1,\, \Jhat_1>
\propto \delta_{J_1-J_2-m,0}\, \delta_{\Jhat_1-\Jhat_2-\mhat,0},
\label{shift}
\eeq
so that the Liouville exponential applied to a highest-weight
state with $J=\Jb$, $\Jhat=\Jhatb$
 only gives states satisfying the same
condition. Thus we may restrict ourselves to the subsector with
zero winding number. We stress this well known fact, since this
will not be true any more in the strong coupling regime.
Let us next
recall some main point of the derivation of the
matrix model results in the present context following
refs.\cite{G5,GR1}. For $C>25$, $Q_L$ is real and $Q_M$ pure
imaginary, so that $\alpha_\pm$ are real. The above formulae are
directly useful. One  represents matter by another copy of the
theory summarized above, now with central charge $c=26-C$, so
that $Q_M$ is its background charge. One constructs local
fields in analogy with Eq.\ref{Lexpdef}:
\beq
e^{\textstyle -(J\alpha'_-+\Jhat\alpha'_+)\Phi'(z, \zb )}=
\sum _{m,\, \mhat}
\Vt'\,^{(J,\,\Jhat)}_{m\,\mhat}(z)\,
\Vtb'\,^{(J,\,\Jhat)}_{m\,\mhat}(\zb).  \>
\label{L'expdef}
\eeq
Symbols pertaining to matter are distinguished by a prime
(or, if more convenient by the index  $M$).
In particular $\Phi'(z, \zb )$ is the matter field  (it
commutes with $\Phi(z, \zb )$), and $\alpha'_\pm$ are the matter
screening charges
\beq
\alpha'_\pm=\mp i\alpha_\mp.
\label{alpha'def}
\eeq
The correct dressing of these operators by gravity is
achieved by considering the vertex
operators
\beq
{\cal W}^{J,\Jhat}\equiv
e^{\textstyle -((-\Jhat-1)\alpha_-+J\alpha_+)\Phi
 -(J\alpha'_-+\Jhat\alpha'_+)\Phi'}
\label{vertexdef}
\eeq
which is an operator of weights $\Delta=\bar \Delta=1$.
In particular for $J=\Jhat=0$, we get the cosmological term
$\exp (\alpha_-\Phi)$. The three-point function was computed
in refs\cite{G5,GR1}. The corresponding product of coupling
constants gives the correct leg factors after drastic
simplifications. After that, one may follow the line of
ref.\cite{dFK} and derive the higher point function.
We will come back to this in the coming section.

\section{The new local fields.}
At this point we turn to the strong coupling regime.
Now $1<C<25$, and $Q_L$,  $Q_M$ are real. The screening charges
$\alpha_\pm$ are complex and related by complex conjugation.
Thus complex weights appear in general. There are  two types of
exceptional cases. The states $|J,\,J>$ (resp. $|-J-1,\,J>$)
 have highest weights which are real and
negative (resp. positive). One could try to work with the
corresponding Liouville exponentials
$\exp[-J(\alpha_--\alpha_+)\Phi]$
(resp $\exp[\bigl((J+1)\alpha_--\alpha_+\bigr)\Phi]$), but this
would be inconsistent, since these operators do not form a closed
set under fusing and braiding. Moreover, as is clear from
Eqs.\ref{Lexpdef}, \ref{shift}, they do not preserve the reality
condition for highest weights just recalled. The basic problem is
that Eq.\ref{Lexpdef} involves the $V$ operators with arbitrary
$m$ $\mhat$, while the reality condition forces us to only use  $V$
operators of the type
\beq
V_{m,\, +}^{(J)}\equiv \Vt_{-m\, m}^{(-J-1,J)},
\quad
V_{m,\, -}^{(J)}\equiv \Vt_{m\, m}^{(J,J)}.
\label{VJpmdef}
\eeq
Now is a good time to recall
 the truncation theorems
which hold for
\beq
C=1+6(s+2),\quad s=0,\pm 1.
\label{Cspecdef}
\eeq
 First  define the physical
Hilbert space
\beq
{\cal H}_{ \hbox{\footnotesize phys}}^{\pm}\equiv
   \bigoplus_{r=0}^{1\mp s}   \bigoplus_{n=-\infty}^\infty
{\cal H}^\pm_{r/2(2\mp s)+n/2}
\label{Hphysdef}
\eeq
where  ${\cal H}^\pm_{J}$  denotes the Verma modules with highest
weights $|\mp(J+1/2)-1/2,\, J>$.
The physical operators $\chi^{(J)}_\pm$
are defined for arbitrary\footnote{
By the symbol
${\cal Z}/(2\pm s)$, we mean the set of numbers $r/(2\pm s)+n$,
with
$r=0$, $\cdots$, $1\pm s$, $n$ integer; $\cal Z$ denotes the set
of all positive or negative integers,
including zero.}  $2J \in {\cal Z}/(2\mp s)$, and
$2J_1 \in {\cal Z}/(2\mp s)$.
 to be such that\footnote{$\cal Z_+$ denotes the set of non
negative
integers.}
\beq
\chi^{(J)}_\pm\>
{\cal P}_{{\cal H}^\pm_{J_1}}
=\sum_{\nu \equiv J+m\in \cal Z_+}
(-1)^{(2\mp s)(2J_1+\nu(\nu+1)/2)}
V_{m,\, \pm}^{(J)} \>{\cal P}_{{\cal H}^\pm_{J_1}},
\label{chidef}
\eeq
where ${\cal P}_{{\cal H}^\pm_{J_1}}$ is the projector
on ${\cal H}^\pm_{J_1}$.
 Denote
 by ${\cal A}^\pm_{\hbox{\footnotesize
phys}}$ the  set   of fields $\chi^{(J)}_\pm $, with
 $2J\in {\cal Z}/(2\mp s)$. The basic
properties  of the special values Eq.\ref{Cspecdef} is the

\noindent TRUNCATION THEOREM:

 For $C=1+6(s+2)$, $ s=0$, $\pm 1$,
and when it acts on
${\cal H} _{ \hbox{\footnotesize phys}}^{+}$
(resp. ${\cal H}\! _{ \hbox{\footnotesize phys}}^{-}$),
the above set ${\cal A}^+_{\hbox{\footnotesize
phys}}$ (resp. ${\cal A}^-_{\hbox{\footnotesize
phys}}$) is closed by braiding and fusion
and only gives states that belong to
${\cal H}_{ \hbox{\footnotesize phys}}^{+}$
(resp. ${\cal H}_{ \hbox{\footnotesize phys}}^{-}$).

Note that the operators $V_{m,\pm}^{(J)}$ themselves
 are not closed by
fusing and braiding, contrary to  the very specific combinations
Eq.\ref{chidef}. The  proof is given in refs.\cite{G3,GR1}. It
follows from a neat  mathematical  property of the quantum
group structure. In general, it  is of the type\cite{G5}
$U_q(sl(2))\odot U_\qhat(sl(2))$ with $h= \pi
\alpha_-^2/2$, $\hhat=\pi \alpha_+^2/2$, so that $h\hhat=\pi^2$.
The two quantum group parameters are thus  related by
duality.  At the special values, one has, in addition
$h+\hhat=s\pi$. Then the q-6j symbols of the two dual quantum
groups become equal up to a sign\cite{GR1}. Using the orthogonality
relation of the
q-6j's this leads to the truncation theorems.

Next we construct local fields out of the chi  fields.
The braiding
of the chi fields is a simple phase. On the unit circle, one has
\beq
\chi^{(J_1)}_\pm \chi^{(J_2)}_\pm =
e^{2i\pi \epsilon (2\mp s)J_1J_2 }
\chi^{(J_2)}_\pm \chi^{(J_1)}_\pm,
\label{chibraid}
\eeq
where $\epsilon=\pm 1$ is fixed by  the ordering of the operator
on the left-hand side in the usual way.   From the spectrum of
the $J$'s, it follows that the phase factor is of the form
$\exp(i\pi N/2(2\mp s))$, where $N\in \cal Z$. Thus, we have
parafermions.  As shown in
ref.\cite{GR1},
 simple products  of the form
$\chi^{(J)}_\pm
\chib^{(\Jb)}_\pm$, with $J-\Jb \in \cal Z$  are local.
 In such a product, the summations over
$m$, and $\mb$ are independent, while the summations over
$m$, $\mhat$, and $\mb$, $\mhatb$ are correlated. Now  we have a
complete reversal of the
weak coupling situation summarized by Eq.\ref{Lexpdef}: the new
fields preserve the reality condition, but {\bf do not preserve
the equality between $J$ and $\Jb$ quantum numbers}.  Thus, as
already stressed in ref.\cite{GR1}, in the strong coupling
regime, we observe a sort of deconfinement of chirality.

\section{The Liouville string}
One may consider two different problems. First, one may build a
full-fledged string theory, by coupling, for instance,
the above  with
$26-C$ free
fields $\vec X$. A  typical string vertex is
of the form
$\exp(i\vec k.\vec X) \chi^{(J)}_+
\chib^{(\Jb)}_+$,
where $\vec k$, $J$, and $\Jb$ are related so that this is
a $1,1$ operator. Here obviously, the restriction to real weight
is instrumental. Moreover, since one  wants the representation of
Virasoro algebra to be unitary, one  only uses the chi+ fields.
This line was already persued with noticable success in
refs.\cite{BG}. However, the N-point functions seem to be beyond
reach at present. Second a simpler problem seems to be tractable,
namely, we may proceed as in the construction of topological
models just recalled. We consider another copy of the present
strongly coupled theory, with central charge $c=26-C$. Since
this gives $c=1+6(-s+2)$, we are also at the special values, and
the truncation theorems applies to matter as well. This ``string
theory'' has no transverse degree of  freedom, and is thus
topological.
The complete dressed vertex operator is now
\beq
\vertex J, \Jb,
=
\chi_+^{(J)} \chib_+^{(\Jb)}\>
\chi'\, ^{(J)}_-  \chib'\, ^{(\Jb)}_-
\label{VertexSdef}
\eeq
As in the weak coupling formula, operators relative to
matter are distinguished by a prime.
The definition of the $\chib$ is similar to the above, with an
important difference. Clearly, the definition Eq.\ref{VJpmdef} of
$V_{m,\, +}^{(J)}$ is not symmetric between $\alpha_+$, and
$\alpha_-$. The truncation theorems also holds if we
interchange the two screening charges. We re-establish some
symmetry between them by taking the other possible definition for
$\chib$, namely, we let
\beq
\Vb_{\mb ,\,+ }^{(\Jb)}\equiv \Vtb_{\mb\,-\mb }^{(\Jb,\,-\Jb-1)},
\quad
V_{\mb ,\, -}^{(\Jb)}\equiv \Vt_{\mb\, \mb}^{(\Jb,\, \Jb)}.
\label{VbJpmdef}
\eeq
Our results will then be  invariant by complex conjugation
provided we exchange $J$'s and $\Jb$'s. Thus left and right movers
are interchanged, which seems  to be a sensible requirement.
For $J=\Jb=0$, we get the new cosmological term
\beq
\vertex 0, 0, =\chi_+^{(0)}(z) \chib_+^{(0)}(\zb).
\label{cosmodef}
\eeq
Thus the area element of the strong coupling regime is
$\chi_+^{(0)}(z) \chib_+^{(0)}(\zb) dz d\zb$. It is factorized
into
a simple product of a single $z$ component by a $\zb$ component.
{}From this expression one may compute the string susceptibility
using the operator version of the DDK argument developed in
ref.\cite{G5} for the weak coupling regime. For this we introduce
the cosmological constant --- so far it was set equal to one.
In ref.\cite{G5}, the weak coupling string
susceptibility was rederived from the following ansatz\footnote{
The previous discussion was actually somewhat different, since
$V$ and $\Vb$ operators were treated differently. This does not
make any difference for the weak coupling regime, but matters
at present.}
\beqa
\left. \Vt_{m\, \mhat}^{(J\, \Jhat)}\right | _{(\mu)}&=&\mu^{J+\Jhat
\alpha_+/\alpha_-}
\mu^{-ip_0/\alpha_-} \Vt_{m\, \mhat}^{(J\, \Jhat )}
\>\mu^{ip_0/\alpha_-} \nnn
\left. \Vtb_{\mb\, \mhatb}^{(\Jb\, \Jhatb )}\right | _{(\mub)}&
=&\mub^{\Jb+\Jhatb
\alpha_+/\alpha_-}
\mub^{-i\pb_0/\alpha_-} \Vtb_{\mb\, \mhatb}^{(\Jb\, \Jhatb )}
\>\mub^{i\pb_0/\alpha_-}
\label{1.10b}
\eeqa
where $p_0$ is the zero-mode momentum of the $\vartheta_L$ free
field.
We take a priori two different parameters $\mu$ and $\mub$. For
the
weak coupling case, the previous discussion is immediately
recovered
with $\mu_c=\mu\mub$ which is the only parameter that counts.
For the strong coupling regime, one substitutes the last equation
into Eq.\ref{chidef}, and its
antichiral couterpart. We have defined the $\chib$ fields so that
taking hermitian
conjugate is equivalent to exchanging $J$'s and $\Jb$'s.
We shall
thus  relate $\mu$ and $\mub$ so that the prefactors are
connected  in
the same way. Taking $\mu$  real, this is realized if
$\mub^{\alpha_+}=\mu^{\alpha_-}$.
Then we have
\beq
\left. \chi_{+}^{(J)}\right|_{(\mu)}
\left.\chib_{+}^{(\Jb)}\right|_{(\mub)} =
\mu^{-2+Q_M(\alpha'_-J+\alpha'_+\Jb)/2}
\mu^{-i(p_0/\alpha_-+\pb_0/\alpha_+)}
\chi_{+}^{(J)} \chib_{+}^{(\Jb)}
\mu^{i(p_0/\alpha_-+\pb_0/\alpha_+)}
\label{chichibmu}
\eeq
For the cosmological term $J=\Jb=0$, and the factor becomes
$\mu^{-2}$.
Thus we conclude that $\mu_c=\mu^2$.  Using the operator
definition of correlators (see refs.\cite{G5,CGR}), and
applying
Eq.\ref{chichibmu},
we get
\beq
\left < \prod_\ell \chi_{+}^{(J_\ell)}
\chib_{+}^{(\Jb_\ell)} \right >_{(\mu_c)}=
\left < \prod_\ell \chi_{+}^{(J_\ell)}
\chib_{+}^{(\Jb_\ell)} \right >
\mu_c^{-\sum_\ell [1-Q_M(\alpha'_-J_\ell+ \alpha'_+
\Jb_\ell)/4]
+(s+2)/2}.
\label{mudep}
\eeq
The last terms emerges when the operators
$\mu^{\pm i(p_0/\alpha_-+\pb_0/\alpha_+)}$ hit the left and the
right vacuum states (of course due to the background charge).
This scaling law is enough to compute\footnote{We only consider
genus zero.}  the string susceptibility.
Consider
\beq
{\cal Z}_{\mu_c}(A)\equiv \left < \delta\left [\int dz d\zb
\left. \chi_{+}^{(0)}\right |_{(\mu_c)}
\left. \chib_{+}^{(0)}\right |_{(\mu_c)} -A\right ]
\right >.
\label{Zmu(A)def}
\eeq
One gets
\beq
\gamma_{\hbox {\scriptsize   str}}=(2-s)/ 2.
\label{1.17}
\eeq
The result is real for $c>1$ ($C<25$), contrary to the
continuation of the
weak-coupling equation
$\gamma_{\hbox {\scriptsize   str}}=2-Q/\alpha_-$.
Explicitly one has
\beq
\left \{
\begin{array}{rccc}
s& c & C & \gamma_{\hbox {\scriptsize   str}} \nnn
1& 7  & 19 & 1/2 \nnn
0& 13 & 13 & 1 \nnn
-1 & 19 & 7 & 3/2 \nnn
\end{array} \right.,\qquad
\left \{
\begin{array}{rccc}
s& c & C & \gamma_{\hbox {\scriptsize   str}} \nnn
2& 1  & 25 & 0 \nnn
-2& 25  & 1 & 2
\end{array} \right.
\label{1.19}
\eeq
The last two are the extreme points of the strong coupling
regime. The values at $c=1$, and $c=25$
 agree with the weak-coupling formula. The result is always
positive,
contrary to the weak-coupling regime. At $c=7$, we find the value
$\gamma_{\hbox {\scriptsize   str}}=1/2$ of branched polymers.

Finally, we have computed  the N-point functions  with one
incoming and N-1 outgoing legs, defined as follows.
First
in general\cite{CGR} the two-point function of two $\Vt$
fields with spins $J_1,\, \Jb_1$, and $J_2,\, \Jb_2$
 vanishes unless $J_1+J_2+1=0$, and $\Jb+\Jb_2+1=0$. Thus
conjugation involves the transformation $J\to -J-1$.
Taking account of the exchange between   $J$ and $\Jhat$,
due to complex conjugation,
yields the following vertex operator for
the conjugate representation:
\beq
\vertexc J, \Jb,
=
\chi_+^{(J)}  \> \chib_+^{(\Jb)} \>
\chi'\,^{(-J-1)}_- \>\chib'\,^{(-\Jb-1)}_-.
\eeq
Thus we have computed the matrix elements
$\left <
\vertexc J_1, \Jb_1,
\vertex J_2, \Jb_2 ,
...
\vertex J_N, \Jb_N,
\right >$. The method is similar to the one developed in ref.\cite{dFK},
with a reshuffling of quantum numbers. In the weak coupling regime,
left and right quantum numbers are kept equal, while the ones  associated
with different  screening charge are chosen independently. In the strong
coupling regime, the situation is reversed: the reality condition
ties the quantum numbers which  differ by the   screening charge, but
the  quantum numbers with different chiralities  become independent.
See refs.\cite{GR3,GR4} for details.
%%%%%%%%%%%%%%%%%%%%%%%%%%%%%%%%%%%%%%%%%%%%%%%%%%%%%%%%%%%
%%%%%%%%%%%%%%%%%%%%%%%%%%%%%%%%%%%%%%%%%%%%%%%

\section{Hints for future developments.}

In the present topological models, both matter and gravity
have a background charge. By construction, the stress-energy
tensor takes the usual free-field form after B\"acklund
transformation.  It is thus  clearly possible to  recombine
the Liouville  B\"acklund field
 $\vartheta_L$, with
its matter counterpart $\vartheta_M$ so that the background
charge appears in one of the free fields only.
For this we  let
\beq
\vartheta_L=-\vec X.\vec \mu_L,\quad \vartheta_M=\vec
X.\vec \mu_M.
\label{X>vartheta}
\eeq
where  $\vec X\equiv
(\varphi,
X)$, and we  introduce the two orthonormalized vectors
\beq
\vec \mu_L=\left ({Q_L\over 2\sqrt{2}},\quad {Q_M\over
2\sqrt{2}}
\right ),\quad
\vec \mu_M=\left ({Q_M\over 2\sqrt{2}},\quad -{Q_L\over
2\sqrt{2}}
\right ).
\label{mudef}
\eeq
Our  conventions   for $\vec X$ coincide with the one of
ref.\citelow{W}, and the present fields are identical
{\bf apart from the zero-mode spectrum.}
 Let $\vec X_0$ be the
center-of-mass position. It is easy to see that
$V^{(J)}_{m,\, \pm }\propto
\exp(im \vec k_\pm.\vec X_0)$, and
$V'\,^{(J)}_{m,\, \pm }\propto
\exp(im\vec k'_\pm .\vec X_0)$, where
$\vec k_-=-iQ_L\vec \mu_L$,
$k_+=Q_M\vec \mu_L$,
$\vec k'_-=-iQ_M\vec \mu_M$, and $
\vec k'_+=Q_L\vec \mu_M$.
Remembering the condition $J+m \in \cal Z_+$ of
eq.\ref{chidef}, we see that
these momenta lie on the lattice generated by the
vectors
\beqa
{-i\over 2\sqrt {2}} \left (1,\, \sqrt{2-s\over 2+s}\right
),&\quad &
{1\over 2\sqrt {2}} \left ( \sqrt{2+s\over 2-s},\, 1\right
),\nnn
{-i\over 2\sqrt {2}} \left (1,\, -\sqrt{2+s\over 2-s}\right
),&\quad &
{1\over 2\sqrt {2}} \left ( \sqrt{2-s\over 2+s},\, -1\right
),
\label{lattice}
\eeqa
which is itself embedded in
a four dimensional space with signature $2,2$. Thus there
may  exist a connection between our topological theories
and  the $N=2$ superstring\footnote{See, e.g.
ref.\cite{OV}.}. Since $Q_M\vec \mu_L/2-Q_L\vec
\mu_M/2
=\left ( 0, \sqrt{2}\right)$, it follows that the momenta of
the $SU(2)$ generators
$\exp\pm i\sqrt{2}X$ belong to the above lattice. Moreover,
$k_{+\, X}=(2-s)/\sqrt{2}$, $k'_{+\, X}=-(2+s)/\sqrt{2}$.
Thus, for integer $s=0$, $\pm 1$ there are points of the lattice
which differ by the momenta $(0, \pm \sqrt{2})$ of the $SU(2)$
generators.
Of course, only the $\vec
X_0$ dependence of the $\chi$ fields  is simple.
In general Eqs.\ref{chidef}, and
\ref{VertexSdef} show that $\vertex J, \Jb, $ is a rather
involved function of $\vec X$ involving momenta of the form
$ m(\vec k_+ +\vec k'_-)$.
Since
that $(\vec k_+ +\vec k'_-)^2=0$, all
our on-shell string states  are massless, and orthogonal to
each other.

Let us turn to a final  remark. The redefinition of the
cosmological term led us to modify the KPZ
formula\cite{KPZ}. On the other hand, in standard studies
of the matrix models  or  KP flows, one first derives
$\gamma_{\hbox {\scriptsize   str}}$ and deduces the
value of the central charge
by  assuming that the KPZ formula holds. In this
way of thinking, one would start from our formula
Eq.\ref{1.17} and apply KPZ, which would lead to
a different value of the central charge, say $d$. It is
easy to see that for $c=1+6(-s+2)$ one gets
$d=1-6(2-s)^2/2s$. This is the value of
a $2, s$ minimal model! What happens is that
in terms of $d$, we have $\gamma_{\hbox {\scriptsize
str}}=(d-1+\sqrt{(d-1)(d-25)})/12$, in contrast with the
KPZ formula $(d-1-\sqrt{(d-1)(d-25)})/12$. Thus the strongly coupled
topological theories  may be given by another branch of $d<1$
theories.

\section{Concluding remarks}
We should probably stress that   no total  conformal spin has been introduced.
Indeed, although, the conformal spin is non zero for the gravity, and matter
components
of our vertex operators separately,
 the  total weight
for the  left and right components are kept equal to one.

One may wonder why the present approach succeeds to break through the $c=1$
barrier, in sharp contrast with the other ones. This may be traced to
the fact that
we first deal with the chiral components of gravity and matter separately
before
reconstructing the vertex operators. This is more painful than the matrix model
approach which directly constructs the expectation values of the dressed matter
operators. However, in this way we have a handle over the way the gravity
quantum
numbers are coupled, and so we may build up vertex operators which change the
gravity chirality. This seems to be the key to the $c=1$ problem, since this
quantum number plays the role of order parameter.

%%%%%%%%%%%%%%%%%%%%%%%%%%%%%%%

\end{document}